\documentclass[aps,prl,twocolumn,superscriptaddress]{revtex4}
\usepackage{graphicx}

\begin{document}
\title{Pinning mode resonance of a Skyrme crystal near Landau level filling factor $\nu$=1}
\author{Han Zhu}
\affiliation{Princeton University, Princeton, NJ 08544, USA}
\affiliation{National High Magnetic Field Laboratory, Tallahassee, FL 32310, USA}
\author{G. Sambandamurthy}
\affiliation{National High Magnetic Field Laboratory, Tallahassee, FL 32310, USA}
\affiliation{Princeton University, Princeton, NJ 08544, USA}
\author{Yong P. Chen}
\affiliation{National High Magnetic Field Laboratory, Tallahassee, FL 32310, USA}
\affiliation{Princeton University, Princeton, NJ 08544, USA}
\author{Pei-Hsun Jiang}
\affiliation{National High Magnetic Field Laboratory, Tallahassee, FL 32310, USA}
\affiliation{Princeton University, Princeton, NJ 08544, USA}
\author{L. W. Engel}
\affiliation{National High Magnetic Field Laboratory, Tallahassee, FL 32310, USA}
\author{D. C. Tsui}
\affiliation{Princeton University, Princeton, NJ 08544, USA}
\author{L. N. Pfeiffer}
\affiliation{Bell Laboratories, Alcatel-Lucent Technologies, Murray Hill, NJ 07974 USA}
\author{K. W. West}
\affiliation{Bell Laboratories, Alcatel-Lucent Technologies, Murray Hill, NJ 07974 USA}

\date{\today}

\begin{abstract}
Microwave pinning-mode resonances found around integer quantum Hall effects, are a signature  of crystallized quasiparticles or holes. Application of in-plane magnetic field to these crystals, increasing the Zeeman energy, has negligible effect on the resonances just below Landau level filling $\nu=2$, but increases the pinning frequencies near  $\nu=1$, particularly for smaller quasiparticle/hole densities. The charge dynamics near $\nu=1$, characteristic of a crystal order, are affected by spin, in a manner consistent with a Skyrme crystal.
\end{abstract}

\pacs{73.43.-f, 73.21.-b, 32.30.Bv}
\maketitle

Spin textures, structures of spin rotating coherently in space, are of importance to a number of different classes of materials \cite{phystoday}. An example of such a spin texture with wide application in descriptions of magnetic order is the skyrmion, a topological defect first used to describe baryons \cite{Skyrme}. Skyrmions can form  arrays which have been considered in a variety of magnetic systems, and which have been seen in bulk material \cite{mnstuff} in magnetic field, in neutron scattering. 
 
Skyrmions are particularly important in two-dimensional electron systems (2DES) with an additional spin or pseudospin degree of freedom \cite{Sondhi,DHLee,Fertig,book}. In these systems, the skyrmion has a unit charge, bundled along with a texture containing multiple flipped spins (or pseudospins) which spread out in space to reduce exchange energy. The skyrmion is also predicted \cite{Fertig} to have a spread-out charge distribution different, for example, from the Landau level (LL) orbitals that would characterize an isolated electron in 2DES.

Skyrmions were identified in 2DES as excitations near Landau level filling $\nu=1$ in experiments \cite{Barrett,Mitrovic,Aifer} that measured the electron spin polarization vs $\nu$, and the presence of skyrmions was also shown  to affect the measured energy gaps in transport \cite{Schmeller,Maude}. The wide applicability of skyrmions in 2DES is attested to also by their presence in layer-index pseudospin of bilayers \cite{Bilayer}, valley pseudospin \cite{Valley} in AlAs, or in predictions involving both valley pseudospin and spin in graphene \cite{Graphene}.

When the skyrmions are sufficiently dilute, they are expected to crystallize, stabilized by Coulomb repulsion. The predicted state has tightly interwoven spin and charge crystal order, and has been of great interest in 2DES \cite{Brey,Cote,Green,Rao,Abolfath,Nazarov,Paredes,Desrat,Gervais,Tracy,Smet,Bayot,Gallais} and has been considered theoretically for  graphene \cite{GrapheneSkCr}. NMR \cite{Desrat,Gervais,Tracy,Smet} and heat capacity \cite{Bayot} experiments in the Skyrme crystal range near $\nu=1$ in 2DES  have focused on the strong coupling of nuclear spins to electron spins, likely via a soft spin wave which was very recently unveiled in Raman scattering \cite{Gallais}. The soft spin mode has been interpreted as arising from XY-spin orientational order \cite{Cote}.

The main result of this paper is that the {\em charge} dynamics of 2DES near $\nu=1$  show characteristics of a Skyrme crystal; we find  these dynamics are  affected by the Zeeman energy and by the charge density in a manner consistent with calculations \cite{Brey,Cote,CotePrivate} done for Skyrme crystals.  Our ability to observe this  demonstrates as well that  the  charge distribution of a skyrmion in a crystal differs from that of an ordinary Landau quasiparticle or hole. The experiments were conducted using broadband microwave conductivity spectroscopy, which has no direct spin sensitivity.  

This microwave spectroscopy has some time ago revealed \cite{Chen} pinned Wigner crystals of the quasiparticles/holes within integer quantum Hall effects. The microwave spectra of these ``integer quantum Hall Wigner crystals", like those of other pinned crystals \cite{SSCReview} in 2DES in high magnetic field,  exhibit a striking resonance, which is understood as a pinning mode resonance \cite{WeakPinning}, due to the collective oscillations of the crystal within the disorder potential that pins it. Pinning modes are present within the integer quantum Hall effects around $\nu=1,2,3$ and $4$ in low-disorder 2DES.  When the exchange energy dominates the  Zeeman energy, skyrmions are expected to be present around $\nu=1$;  it is well established \cite{Schmeller,highNu} that near other integer $\nu$ under any experimentally accessible conditions in 2DES, skyrmions do not form.

This paper addresses the nature of the crystal around $\nu=1$, and shows that the pinning mode near $\nu=1$ is affected by skyrmion formation. These effects become clear through systematic study of the dependence of the pinning mode on in-plane field, $B_{\parallel}$, which at fixed $\nu$  reduces  the skyrmion spin \cite{Sondhi,DHLee,Fertig,book}. $B_{\parallel}$ has essentially no effect on the crystal just below $\nu=2$, expected to be a Wigner crystal of single Landau quasiholes.   In contrast, $B_{\parallel}$ increases the pinning frequency of the crystal near $\nu=1$. In addition, the skyrmion density (the same as the charge density) in the crystal increases for $\nu$ farther from exactly 1, and is also large for higher electron density 2DES. We find $B_{\parallel}$ increases the pinning mode frequency only for small enough skyrmion density, or wide enough skyrmion separation. This shows that spin size of skyrmions is affected by their mutual proximity \cite{Cote}, an illustration of the intertwined nature of charge and spin in this state.

We used 2DES's from two high quality wafers of GaAs/AlGaAs/GaAs quantum wells. Sample A is a 30 nm quantum well (\#5-20-05.1), with the electron density $n=2.7\times10^{11}/\ \text{cm}^2$, and a low temperature mobility $\mu=27\times10^6\ \text{cm}^2/\text{Vs}$. Sample B is a 50 nm quantum well (\#7-20-99.1), with a lower $n=1.1\times10^{11}/\ \text{cm}^2$ and $\mu=15\times10^6\ \text{cm}^2/\text{Vs}$.

The microwave spectroscopy technique is similar to those reported earlier \cite{SSCReview,Chen}. As illustrated in Fig. 1(a), we deposited on the sample surface a metal-film coplanar waveguide, with length $l=4\sim28$ mm and slot width $W=30\sim80\ \rm{\mu m}$, impedance matched to $Z_0=50\ \Omega$. The microwaves couple capacitively to the 2DES. From the absorption $P$ of the signal by the 2DES, we can calculate the real part of its diagonal conductivity $\text{Re}\left[\sigma_{\text{xx}}\left(f\right)\right]=\left(W/2lZ_0\right)\text{ln}P$.

To tilt the sample in magnetic field, we designed a rotator with low-reflection broadband microwave connections  through flexible microstrips. The tilting angle $\theta$ between the sample normal axis and the field direction is calculated from the total fields $B_{\rm{tot}}$ of prominent integer quantum Hall states. The perpendicular field component $B_{\perp}=B_{\rm{tot}}\cos\theta$. The measurements were done at the low power limit and the sample temperature was 40 mK.

\begin{figure}
		\includegraphics[width=.45\textwidth]{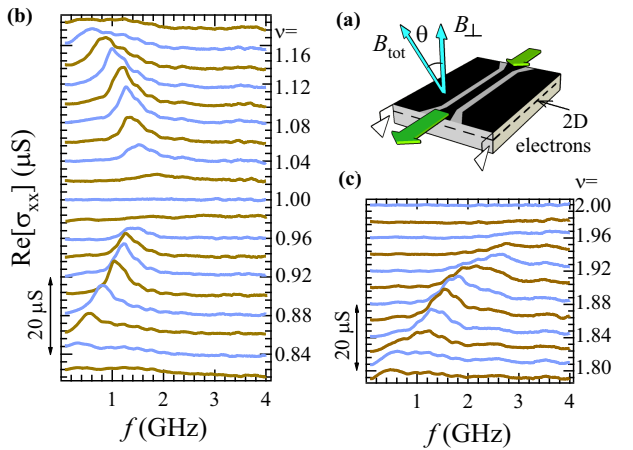}
\caption{\label{Fig1}(color online). (a) An illustration of the coplanar waveguide. (b) From sample A, at $\theta=0^{\rm{o}}$, the real conductivity spectra at various $\nu$,  increasing from $0.82$ to $1.18$ in steps of $0.02$. (c) Also taken at $\theta=0^{\rm{o}}$, the real conductivity spectra at $\nu$ from $1.78$ to $2.00$, increasing in steps of $0.02$.}
\end{figure}

For sample A, Fig. 1 shows a set of conductivity spectra (b) around $\nu=1$ and (c) just below $\nu=2$. The resonances in the spectra, which flatten out for temperatures above 150 mK, are understood as the pinning modes of the crystallized quasiparticles/holes, as in Ref. \cite{Chen}. In the following, for brevity, we shall refer to $\nu$ just below $2$ which exhibit the resonance as $\nu=2^-$, and the resonant $\nu$ ranges  just above (below) $1$ as $\nu=1^+$ ($1^-$). In both Fig. 1(b) and (c), the resonance peak frequency $f_{\rm{pk}}$ decreases with increasing partial filling factor $\nu^*=|\nu-$nearest integer$|$. This is consistent with the picture of a weakly pinned crystal \cite{WeakPinning}, because higher $\nu^*$ means denser quasiparticles/holes and stronger interaction, and thus a weaker role of disorder. Comparing Fig. 1 (b) and (c), we notice that, close to integer $\nu$, the $\nu=2^-$ resonances have a stronger $\nu^*$-dependence of $f_{\rm{pk}}$ than the $\nu=1^+$ and $1^-$ resonances.

\begin{figure}
		\includegraphics[width=.45\textwidth]{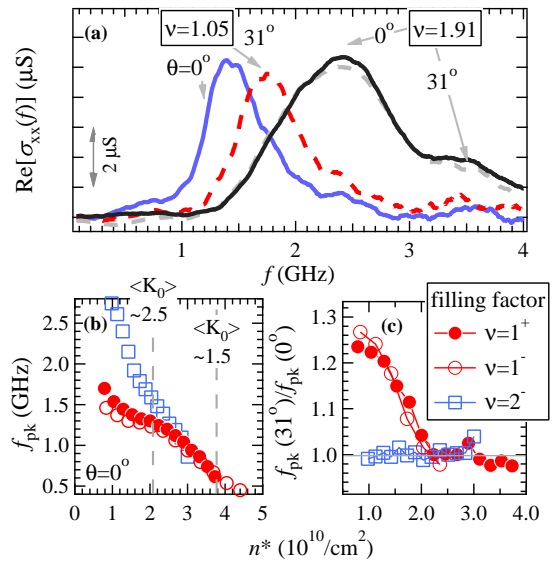}
\caption{\label{Fig2}(color online). (a) For sample A, the real conductivity spectra at $\nu=1.05$ and $\nu=1.91$, measured at $\theta=0^{\rm{o}}$ and $31^{\rm{o}}$. (b) For sample A at $\theta=0^{\rm{o}}$, the resonance peak frequencies $f_{\rm{pk}}$ as a function of quasiparticle density $n^*$ for the $\nu=2^-$ (open squares), $\nu=1^+$ (solid circles), and $\nu=1^-$ (open circles) crystals. The skyrmion sizes, $\left\langle K_0\right\rangle$, estimated by Ref. \cite{Cote}, are marked in the graph. (3) The relative change in $f_{\rm{pk}}$ from $\theta=0^{\rm{o}}$ to $31^{\rm{o}}$.}
\end{figure}

Still for sample A, we repeat the measurement in a tilted magnetic field. We fix $B_{\perp}$ and hence $\nu$, and increase $B_{\rm{tot}}$ by a factor of $1/\cos\theta$. The Zeeman energy also increases, since it is proportional to $B_{\rm{tot}}$. Fig. 2 (a) shows the spectra at $\nu=1.05$ and $1.91$, each taken at $\theta=0^{\rm{o}}$ and $31^{\text{o}}$. At these two $\nu$, the quasi-particles/holes are within the same orbital LL ($N=0$), and have equal density $n^*\sim 1.3 \times10^{10}/\text{cm}^2$, calculated as $(\nu^*/\nu)n$. At $\theta=0^{\text{o}}$, the $\nu=1.05$ resonance peak is at about half the frequency of the $\nu=1.91$ resonance. Tilting to $31^{\text{o}}$ has negligible effect on the $\nu=1.91$ resonance, but shifts the $\nu=1.05$ resonance to higher frequency. The $\nu=0.95$ resonance, not shown here, behaves almost identically to the $\nu=1.05$ resonance.

Panels (b) and (c) in Fig. 2 present the effect of $B_{\parallel}$ for $\nu=1^+$, $1^-$ and $2^-$. Fig. 2(b) uses the $\theta=0^{\rm{o}}$ data in Fig. 1, and plots $f_{\text{pk}}$ vs quisiparticle/hole density $n^*$, for all three crystal phases: $\nu=2^-$, $\nu=1^+$, and $\nu=1^-$. All the three curves show $f_{\rm{pk}}$ decreases as $n^*$ increases, as we have seen in Fig. 1. The $\nu=1^+$ and $\nu=1^-$ curves agree well, indicating good symmetry between quasiparticles and quasiholes, or, as we shall see, skyrmions and anti-skyrmions. The $\nu=1^+$ and $1^-$ curves also overlap with the $\nu=2^-$ curve for high $n^*$. But for low $n^*$ ($<2.5\times10^{10}/\rm{cm}^2$), the $\nu=1^+$ and $1^-$ resonances have lower $f_{\rm{pk}}$ than the $\nu=2^-$ resonance for the same $n^*$.

Fig. 2(c) plots the ratio, $f_{\text{pk}}(31^{\rm{o}})/f_{\text{pk}}(0^{\rm{o}})$, of $\theta=31^{\rm{o}}$ and $0^{\rm{o}}$ $f_{\rm{pk}}$. $f_{\text{pk}}$ increases with tilting only for the $\nu=1^+$ and $1^-$ resonances with low $n^*$ ($<2.5\times 10 ^{10} /\rm{cm}^2$). The two curves of $f_{\text{pk}}$ vs $n^*$, for $\nu=1^+$ and $1^-$, agree with each other, indicating a still good particle-hole symmetry. Tilting has negligible effect on the $\nu=2^-$ resonances, or the $\nu=1^+$ and $1^-$ resonances with high $n^*$.

\begin{figure}
		\includegraphics[width=0.34\textwidth]{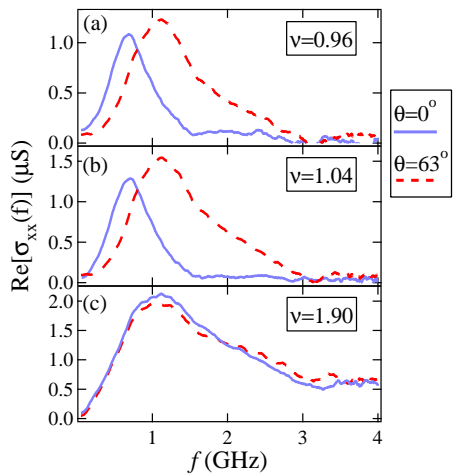}
\caption{\label{Fig3}(color online). The conductivity spectra of sample B, taken with $\theta=0^{\rm{o}}$ (solid curve) and $\theta=63^{\rm{o}}$ (dashed curve), at (a) $\nu=0.96$, (b) $\nu=1.04$, and (c) $\nu=1.90$.}
\end{figure}

Sample B has lower $n$ than sample A, and as discussed below, is predicted to be capable of supporting larger skyrmions. Near integer $\nu$, sample B also displays resonances in the real conductivity spectra. Fig. 3 shows the resonances at three typical $\nu$: (a) $0.96$, (b) $1.04$, and (c) $1.90$, each taken at $\theta=0^{\rm{o}}$ and $\theta=63^{\rm{o}}$. Tilting shifts the $\nu=0.96$ and $\nu=1.04$ resonances to higher frequencies, in clear contrast to the negligible effect it has on the $\nu=1.90$ resonance.

\begin{figure}
		\includegraphics[width=0.45\textwidth]{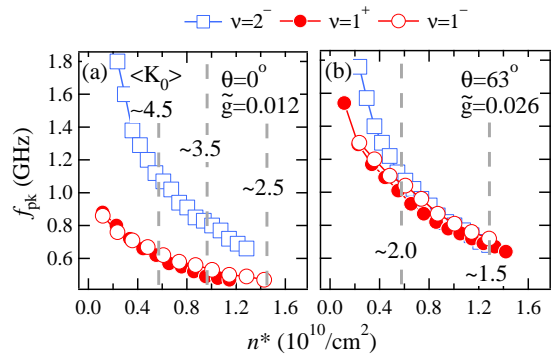}
\caption{\label{Fig2}(color online). From sample B, $f_{\text{pk}}$ as a function of $n^*$ for the $\nu=2^-$ (open squares), $\nu=1^+$ (solid circles), and $\nu=1^-$ (open circles) resonances, at (a) $\theta=0^{\rm{o}}$ and (b) $\theta=63^{\rm{o}}$. The skyrmion sizes, $\left\langle K_0\right\rangle$, estimated by Ref. \cite{Cote}, are marked in the graph.}
\end{figure}

For sample B, at $\theta=0^{\rm{o}}$ and $63^{\rm{o}}$ respectively, Fig. 4 (a) and (b) show $f_{\rm{pk}}$ vs $n^*$ for $\nu=1^+$, $1^-$ and $2^-$. Here we also see that $f_{\text{pk}}$ decreases as $n^*$ increases, consistent with a weak pinning picture. The $\nu=1^+$ and $1^-$ resonances show good agreement with each other , but have lower $f_{\rm{pk}}$ than the $\nu=2^-$ resonances, with the difference greater for smaller $n^*$. %
Comparing Fig. 4 (a) at $\theta=0^{\rm{o}}$ and (b) at $63^{\rm{o}}$, we see $B_{\parallel}$ has negligible effect on the $\nu=2^-$ resonances but increases the frequencies of the $\nu=1^+$ and $1^-$ resonances.


By analogy to the Wigner crystal pinning modes found at very low $\nu$  ($\lesssim 1/5$) \cite{SSCReview}, the observed   resonances at $\nu$ just away from 1 and other integers are clear signatures of crystallized quasiparticles/holes \cite{Chen}. Here we show that the dependence on $B_{\parallel}$ indicates a distinct difference of the crystals of  $\nu=1^+$ and $1^-$ from that of  $\nu=2^-$, and provides evidence for a Skyrme crystal near $\nu=1$.

The effect of $B_{\parallel}$ cannot be explained without invoking the spin degree of freedom. This is clear from the negligible effect of $B_{\parallel}$, except for the crystals with low $n^*$ near $\nu=1$. The crystals studied here, for $\nu=2^-$, $1^+$ and $1^-$, all belong to the lowest orbital LL. If the effect of $B_{\parallel}$ were due to the orbital wave function, it would be unlikely that the same effect would be completely absent for the $\nu=2^-$ crystal and the $\nu=1^+$ and $1^-$ crystals with high $n^*$ \footnote{$B_{\parallel}$ is found to have a very different effect on the higher orbital LL's, which shall be presented in another publication.}.

The effect of $B_{\parallel}$ near $\nu=1$ finds a natural explanation in a predicted crystal of skyrmions \cite{Brey,Green,Cote,Rao,Abolfath,Nazarov,Paredes}. The skyrmion size, $\left\langle K_0\right\rangle$, denotes the number of flipped spins (relative to the maximally spin-polarized state with the same charge) \cite{Fertig,Cote}. Two key parameters control $\left\langle K_0\right\rangle$. One is the Zeeman/Coulomb energy ratio \cite{Fertig}, $\tilde{g}=g\mu_{B}B_{\text{tot}}/\left(e^2/\epsilon l_{0}\right)$, where $\left|g\right|\approx0.44$ is the $g$ factor in GaAs, $\mu_{B}$ the Bohr magneton, $\epsilon\approx13$ the dielectric constant of GaAs, and $l_{0}=\sqrt{\hbar/eB_{\perp}}$ the magnetic length. Larger $\tilde{g}$ makes flipping spins more costly, and favors smaller $\left\langle K_0\right\rangle$. The second parameter is $n^*$: increasing $n^*$ by tuning $\nu$ away from 1 brings skyrmions closer and limits $\left\langle K_0\right\rangle$. Ref. \cite{Cote} calculated $\left\langle K_0\right\rangle$,  predicting the reduction of $\left\langle K_0\right\rangle$ on increasing  $\tilde{g}$ or $n^*$.%

For our samples, A has $\tilde{g}\sim0.019$ at $\nu=1$ in perpendicular field, and B with a lower $n$ has a smaller $\tilde{g}\sim0.012$, because $\tilde{g}\sim \sqrt{n}$. Upon tilting, $\tilde{g}$ increases by a factor of $1/\cos\theta$, the same as $B_{\rm{tot}}$. Based on Ref. \cite{Cote}, the calculated $\left\langle K_0\right\rangle$ for different $\tilde{g}$ and $n^*$ are marked in Fig. 2(b) and Fig. 4.

The data taken near $\nu=1$ in both samples show a strikingly consistent correlation between $f_{\rm{pk}}$ and  the calculated $\left\langle K_0\right\rangle$. The correlation also holds for intermediate $\theta$, for example $39^{\rm{o}}$ for sample B, not just the $\theta$ shown in the graphs. The negligible effect $B_{\parallel}$ has on $\nu=2^-$ serves as a control, since the absence of skyrmions there is well established. For the same $n^*$, when $\left\langle K_0\right\rangle$ is small, the pinning mode frequencies $f_{\rm{pk}}$ of the $\nu=1^+$ and $1^-$ crystals are close to that of the $\nu=2^-$ crystal. The point where $\nu=1^+$ and $1^-$ crystals start to have lower $f_{\rm{pk}}$ than the $\nu=2^-$ crystal is always at $\left\langle K_0\right\rangle\sim2$, and the difference widens as $\left\langle K_0\right\rangle$ increases. The separation of the curves occurs consistenly at $\left\langle K_0\right\rangle\sim2$, even though this condition occurs  at different $n^*$ when $\tilde{g}$ changes.

At fixed $n^*$, decreasing $B_{\parallel}$ produces larger predicted $\left\langle K_0\right\rangle$, and we find a lower $f_{\rm{pk}}$. This suggests a crystal formed by larger skyrmions is more weakly pinned. This effect can be due to two possible mechanisms which may both be operational:  one   from skyrmion-disorder interaction and the other from skyrmion-skyrmion interaction. Larger skyrmions average disorder over a larger area, resulting in an effectively weaker disorder, and hence a lower $f_{\rm{pk}}$. $f_{\rm{pk}}$ is also sensitive to the shear modulus, which is determined by the interaction between the skyrmions. When the skyrmions grow bigger and begin to overlap, the inter-skyrmion repulsion is expected to get stronger and to increase the shear modulus of the crystal \cite{CotePrivate}. And within the weak pinning picture \cite{WeakPinning}, a stiffer crystal has lower $f_{\rm{pk}}$, because it is more difficult for the crystal to deform to take advantage of the disorder landscape.

As $\nu$ departs from 1, a phase transition is predicted, from a triangular-lattice ferromagnet of spin helicity to a square-lattice antiferromagnet \cite{Brey,Green,Cote,Rao,Abolfath,Nazarov,Paredes}, in favor of lower exchange energy at a cost in the Madelung energy. From calculations in Ref. \cite{Cote}, this is predicted to happen roughly at $\left\langle K_0\right\rangle\sim2.5$ in Fig. 2(b), $\sim 5$ in Fig. 4(a) and $\sim1.5$ in Fig. 4(b). In our measurement we do not see any abrupt jump in the pinning frequencies. One possible explanation is that sample disorder smears the phase boundary.

To summarize, pinning mode resonances observed around integer  $ \nu$  provide clear evidence of crystal phases of the quasiparticles. The pinning modes near $\nu=1$ appear to be consistently determined by the  predicted skyrmion sizes, indicating  the crystal around that filling is formed of skyrmions at least for small $n^*$. As $B_{\parallel}$ causes the predicted skyrmions to shrink, we observe upshift of the pinning frequency, which saturates as skyrmions approach single spin flips. The dependence of the pinning on the spin  is a consequence of  the coupled charge and spin degrees of freedom for crystallized skyrmions.

We thank R. C\^ot\'e, H. A. Fertig and Kun Yang for helpful discussions, and we thank G. Jones, J. Park, T. Murphy and E. Palm for experimental assistance. This work was supported by DOE Grant Nos. DE-FG21-98-ER45683 at Princeton, DE-FG02-05-ER46212 at NHMFL. NHMFL is supported by NSF Cooperative Agreement No. DMR-0084173, the State of Florida and the DOE.

\end{document}